# The Influence of High-Energy Lithium Ion Irradiation on Electrical Characteristics of Silicon and GaAs Solar Cells

B. Jayashree, Ramani, M.C.Radhakrishna, Anil Agrawal, Saif Ahmad Khan, and A. Meulenberg

*Abstract*— **Space-grade Si and GaAs solar cells were irradiated with 15 & 40 MeV Li ions. Illuminated (AM0 condition) and unilluminated I-V curves reveal that the effect of high-energy Li ion irradiation has produced similar effects to that of proton irradiation. However, an additional, and different, defect mechanism is suggested to dominate in the heavier-ion results. Comparison is made with proton-irradiated solar-cell work and with non-ionizing energy-loss (NIEL) radiation-damage models.**

## I. INTRODUCTION

Electron and proton damage analysis of solar cells is important for predicting their response to radiation environments in space. Electrons and protons with a wide range of energies dominate the space-radiation environment. Generally the radiation tolerance of space solar cells has been evaluated using 1 MeV electrons with fluence up to $1 \times 10^{16}$ e/cm$^2$ and 10 MeV protons with fluence up to $1 \times 10^{13}$ p/cm$^2$ [1]. For solar cell types that have shown reliable performance in space, 1 MeV electron radiation-damage studies of the electrical properties, such as short-circuit current (Isc), open-circuit voltage (Voc), and maximum-power output (Pmax), are generally sufficient and they constitute "acceptance" tests. Acceptance tests are generally based on a set of criteria that characterize individual cell performance relative to the standard performance of a cell type for a particular radiation environment [2] that is readily available to the community. These tests are typically performed on a statistically significant number of samples to represent a qualified process for a particular program.

If new cell types are introduced, or if different radiation environments are anticipated, then more complete testing, or different type testing is required. Such testing is classified as "qualification" testing with the goal of determining the criteria for the acceptance testing of an individual program or procurement. In some cases, even a "prequalification" test program is required, since there may be some doubt as to whether a specific type of testing is appropriate to a given process or environment. This prequalification testing may be better classified as R&D rather than engineering testing. It requires a different type of analysis and assurance that no "fatal" flaws are being overlooked. This is more important than a statistically-significant number of samples.

In a supplementary approach and a refinement to standard radiation testing and radiation-damage prediction of solar cells, work is continuing on a methodology to predict degradation based on models of non-ionizing energy loss (NIEL) for different particle types and energy [3]. Since the NIEL of an incident energetic particle is the primary source of displacement damage in a junction semiconductor and little permanent damage is caused by the energy lost by a particle from ionization, this is a sound approach. This approach is distinct from the extensive models that have been developed to predict and test semiconductor devices which incorporate non-conductors, which could trap ionization products (electrons, ions, and holes) thereby altering the device performance.

Recognizing that different devices are affected differently by radiation, absolute damage characteristics are not required, (or sometimes even attempted) in this NIEL process. However, once a damage/performance characteristic of a device is determined in an experiment, the NIEL model can be employed to predict degradation from other types and energies of laboratory particles.

The NIEL approach has a double benefit. First, as stated above, it can provide a tool for determining the damage equivalence of different laboratory-radiation sources. The second benefit is similar to the earlier development of the

Manuscript received July 10, 2006. Bangalore University, Bangalore, India. B. Jayashree was provided financial support by University Grants Commission, New Delhi, India. A. Meulenberg was supported by the Science for Humanity Trust, Bangalore, India.

B. Jayashree is with the Dept. of Physics, Bangalore University, Jnana Bharathi campus, Bangalore-560056 India. (Phone: +91-8026720780, fax: +91-80-23219295, email: jaya_mdev@yahoo.co.in )

Ramani is with the Department of Physics, Bangalore University, Jnana Bharathi campus, Bangalore-560056 India. (email: ramanirm@hotmail.com)

M.C.Radhakrishna is with the Department of Physics, Bangalore University, Jnana Bharathi campus, Bangalore-560056 India. (email: mcradhakrishna@yahoo.co.in)

Anil Agrawal is with the Solar Panels Division, Power Systems Group, ISRO Satellite Centre, Vimanapura, Bangalore-560017, India. (email: daa@isac.gov.in)

Saif Ahmad Khan is with the Inter-university Accelerator centre, Arun Asaf Ali Marg, P.O.10502, New Delhi-110067, India, (email: saif@nsc.ernet.in)

A. Meulenberg is Visiting Scientist - Department of Instrumentation, Indian Institute of Science, Bangalore 560012, India (email: mules333@gmail.com ) equivalent-fluence [2] models needed for acceptance tests and critical for determining a range of acceptance tests for solar cells (and other minority-carrier devices) to be used in space. These models are based on an attempt to determine the fluence of a given laboratory source required to give a damage equivalence for a certain period of time in a specific space environment.

The NIEL approach is able to take advantage of extensive codes previously developed for other purposes and is a powerful tool for comparing damage from space and laboratory sources. Nevertheless, despite the general acceptance of the NIEL approach, there are problems that hinder its widespread use [3]. It is expected that the present work will contribute to the data base required for improvement of the model and its understanding.

The emphasis of the present work is on laboratory simulation with different radiation rather than on space-radiation effects or on application of the model to new materials. The non-ionizing energy loss is related to displacement damage and, therefore, to the energy spectrum of the primary knock-on atom (PKA). Simulation of this type of damage is thus of primary importance.

Lithium ions have several advantages over protons as tools with which to study the details of PKA's. Firstly, by being triply charged, they are able to be accelerated to much higher energy than that of protons in the same machine. Secondly, being 7 (or 6) times more massive than protons, Li ions can deliver more energy to PKA's and therefore can cover an even greater range of PKA energies for a given accelerator. Thirdly, still being a penetrating particle, at reasonable energies the non-directly-colliding Li ions can penetrate through the active volume of modern solar cells. This eliminates a conflicting defect type (a dead or undoped region) that is seldom observed in space, but is often a problem in laboratory simulations with monoenergetic particles [4].

An attempt is made here to establish a relation between the radiation-induced damage from protons and Li ions, to provide input to the radiation models, available and being developed, and to establish a basis for a lithium equivalent to proton irradiation (at least for a defect type that may become critical for future cells and for simulation of specific orbital environments).

## II. EXPERIMENTAL DETAILS

The samples for this program are industry-standard space–grade Si and GaAs/Ge solar cells (Table I). As such, they have been well characterized over the years [1], [5]. Since this work is seeking details of a particular damage mechanism, statistics have little meaning. The important thing is to provide a set of well-characterized cells with damage over a significant range of particle fluences. With such a set, we will be able to separate out the defects from PKA's that affect the solar-cell performance.

The GaAs/Ge cells (as received) were covered; therefore, they had to be "delidded" prior to irradiation. This process was not easy and some cells were destroyed in the process.

High energy (15 & 40 MeV) Li irradiation at room temperature in vacuum was carried out using 16 MV Pelletron Accelerator [6] at the Inter-University Accelerator Center (IUAC), New Delhi, India. The solar cells were irradiated with an ion fluence ranging from $10^{10}$ ions cm$^{-2}$ to 5x10$^{12}$ ions

TABLE I
SOLAR CELL SPECIFICATION

| PARAMETERS | GaAs/Ge CELL | Si CELL |
|---|---|---|
| Manufacturer | Tecstar | RWE Space Solar Power GMBH |
| Configuration | P/N type | N/P type |
| specifics | Sb added for lattice matching | BSR, dual ARC (no BSF) |
| Initial efficiency (nominal) | 18.5% | 13.0% |
| Base resistively | 0.01 Ωcm | 2 Ωcm |
| Base doping levels | 2x10$^{17}$/cm$^3$ | ~8x10$^{15}$/cm$^3$ |
| Size | 2x4 cm$^2$ | 3.5x7.5cm$^2$ |
| Nominal thickness | 145 μm | 200 μm |
| Nominal active thickness | 5 μm | 200 μm |
| Preparation Technique | MOVPE | Diffused Junction |

cm$^{-2}$ (Table II)[a] at 300K in an experimental chamber of diameter 1.5 m maintained at 10$^{-7}$ mbar of vacuum.

The ion beam was scanned over the cell (in 1.5 x 1.5 cm sections) with a magnetic scanner to obtain uniform fluence over the cells. The beam current was typically in the range of 1-3 pnA (particle nano-Ampere) for both beam energies (Li$^{++}$ for 15 MeV and Li$^{+++}$ for 40 MeV). While the Li charge differs for the two beam energies, the charge establishes equilibrium almost immediately upon contact with the target. The area-integrated flux was of the order of 2 x 10$^9$/cm$^2$ sec. The instantaneous flux density was much greater because of the much-smaller beam size (~2 mm diam.). Due to the nature of the defect mechanism being sought, a maximum flux density of up to ~6x10$^{11}$/cm$^2$ sec was not considered excessive.

TABLE II
BEAM FLUENCE AND ENERGY

| Si Cells | GaAs Cells | GaAs Cells |
|---|---|---|
| 40 MeV Fluence (Ions/cm$^2$) | | 15 MeV Fluence (Ions/cm$^2$) |
| 5X10$^{10}$ | 1X10$^{10}$ | |
| 1X10$^{11}$ | 1X10$^{11}$ | 1X10$^{11}$ |
| 5X10$^{11}$ | | 5X10$^{11}$ |
| 1X10$^{12}$ | 1X10$^{12}$ | 1X10$^{12}$ |
| 5X10$^{12}$ | 5X10$^{12}$ | 5X10$^{12}$ |

After irradiation, all of the cells were allowed to recover at room temperature for 6 days before being measured.

The air mass zero (AM0) properties of the cells before and after irradiation were measured (at 25.7°C) using the Indian Space Research Organization (ISRO) X-25 solar simulator with illumination intensity set to 135.3 mWcm$^{-2}$ and the dark properties of the cells were measured using a Keithley 2420, 3A source meter.

---
[a] The 1x10$^{11}$ 40 MeV GaAs cell was damaged post-rad; but, before measurement.

The day-to-day and run-to-run measurement reproducibility for the AM0 measurements was good. A standard deviation of $\sigma = \sim 0.1\%$ in Isc indicated good stability and reproducibility in the X-25 intensity. The temperature control was indicated by a $\sigma = \sim 0.3\%$ (fraction of a mV) in Voc. The reproducibility in Pmax was not quite as good as that of Isc and Voc. This was attributed to a problem in the contacts. Occasional values were low and it was observed, in retrospect, that there might be a sequence of low values that was terminated when the contacts were cleaned. In most cases, the low values could be rejected from the data which included multiple measurements for each cell. In the few cases where this procedure was not adequate, the recorded increase in series resistance does not present a significant problem.

### III. AM0 MEASUREMENT RESULTS

Figs. 1 and 2 show the AM0 I-V curves for the Si and GaAs solar cells irradiated with 40 MeV Li ions. In the silicon cells, the knee (at Pmax) begins to "soften" at $5 \times 10^{11}$ Li /cm$^2$ and the IV characteristic begins to fall apart above that fluence level. It is not expected that space solar cells would be operated in this region nor would standard IV analysis would be effective for the results above this fluence level. In the GaAs cells, the knee has softened noticeably at $5 \times 10^{12}$ Li /cm$^2$; nevertheless, the cell IV-characteristic curve is still significantly better than that for the silicon cell at 1/10 the fluence.

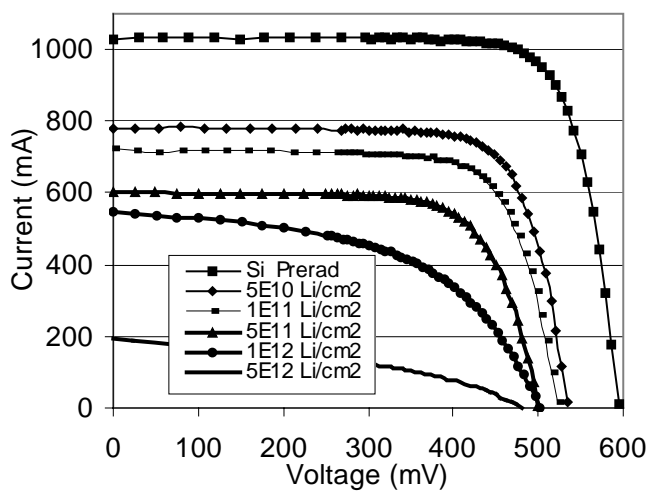

Fig.1. I-V curves for Si solar cells irradiated with 40 MeV Li ions.

An unusual feature of the silicon data set is the fact that, when the knee really begins to collapse (at $1 \times 10^{12}$), the voltage does not. This is explained by series resistance Rs resulting from undoping of the base ($\sim 8 \times 10^{15}$ B/cm$^3$). The series resistance, at high currents, skews the IV curve towards the left since the junction voltage Vj (Vj = V ± IRs in (1) [5], where the sign depends on dark or illuminated IV measurement) is no longer equal to the applied voltage V. Since Isc is defined as the cell current I at Vj = 0, it is found at V < 0 for non-zero Rs. The slope of the IV curves near V = 0 is a result of growth in the shunt resistance Rsh. This resistance allows a shunt current (Ish = Vj / Rsh) to flow across the junction without contributing to the output power. The shunt current grows with applied voltage and is subtracted from Isc. Therefore, with combined increase in Rs and decrease in Rsh, the measurement of Isc at V = 0 can be significantly lower than a "true" Isc value.

$$I = I_{sc} - I_0 \left[ \exp\{q(V \pm IR_s)/nKT\} - 1 \right] - (V \pm IR_s)/Rsh \quad (1)$$

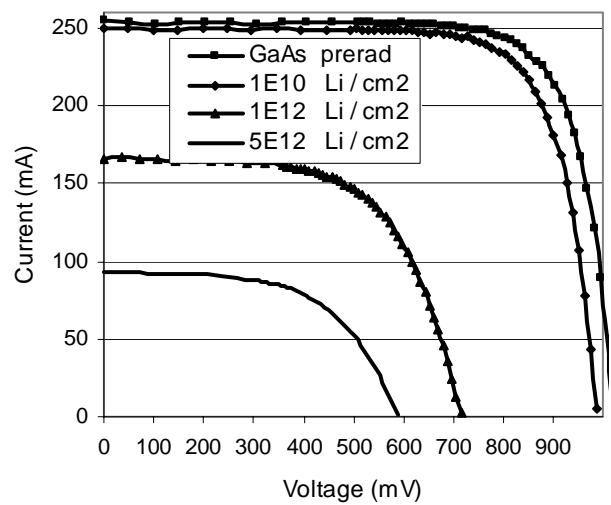

Fig.2. I-V curves for GaAs cells irradiated with 40 MeV Li ions.

The true Isc and the Voc do not degrade much from a fluence of $5 \times 10^{11}$ to $1 \times 10^{12}$ because the Fermi level in the base begins to move towards mid-gap at this fluence and thus reduces the charge (and therefore the minority-carrier-capture cross section) of shallow defect levels in the band gap. In Fig. 2, this severe undoping behavior does not occur in the more heavily-doped ($2 \times 10^{17}$/cm$^3$) GaAs/Ge cells, even at the maximum fluence. Nevertheless, even at $1 \times 10^{12}$, a decided change in slope beyond the knee (i.e., at higher voltages) is observed. This would appear to be a result of an increase in series resistance.

Fig. 3, with 15 MeV Li ion data, provides confirmation of the nature of the 40 MeV Li ion damage to GaAs in Fig. 2. The results appear to be well behaved except for the data at $1 \times 10^{12}$ where the curve should fall closer to that for the $5 \times 10^{11}$ results rather than mid way between those of the $5 \times 10^{11}$ and $5 \times 10^{12}$ results. Figs. 4 and 5 display the degradation rates of electrical parameters of the Si and GaAs solar cells (of Figs. 1 and 2) irradiated with 40 MeV Li ions. Since the data at any given fluence represent a single cell, the fit of data to the trendlines drawn is not expected to be comparable to the measurement precision. The trendlines for the relative degradation of IV parameters, such as Pmax (Pmax /P$_0$), are fitted curves based on the logarithm of the ratio of irradiation fluence φ to a "fitting" fluence φ$_o$, (2), and are representative of curve shapes obtained with large sets of irradiated cells [5]. These trendlines will be used for comparison of the present work with prior proton irradiations and in degradation vs. NIEL plots.

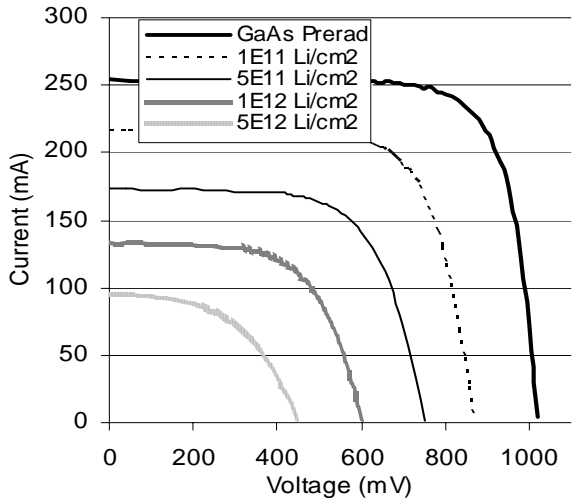

Fig. 3. The I-V curve for GaAs cells irradiated with 15 MeV Li ions.

$$P\max / P_0 = 1 - C \log(1 + \phi / \phi_0) \quad (2)$$

In notable cases in Fig. 4, a data point has been ignored in drawing the trendline. The Pmax point for the silicon cell at $5 \times 10^{10}$ is high because the unirradiated cell had a weak fill factor (determined by repeated measurements). Since Fig. 4 displays relative measurements, and irradiation of silicon increases the defect levels that primarily contribute to the "bulk" component (n = 1) of the IV characteristic, the fill-factor actually improved with initial radiation. Therefore, this data point is expected to be high and was not allowed to influence the trendline.

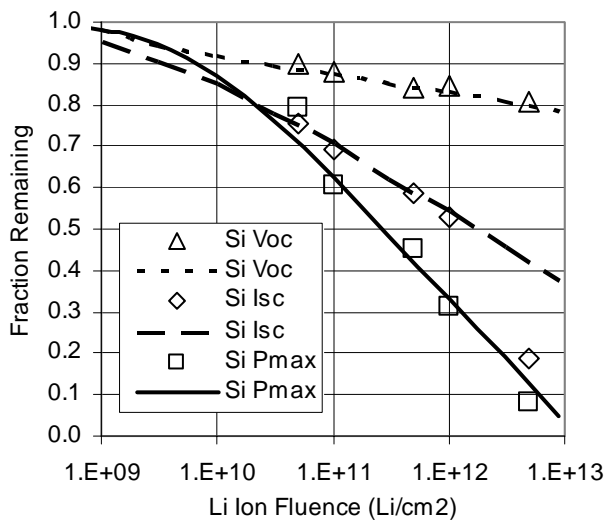

Fig. 4. IV parameter degradation rates (trendlines and data) of Si solar cells irradiated with 40 MeV Li ions.

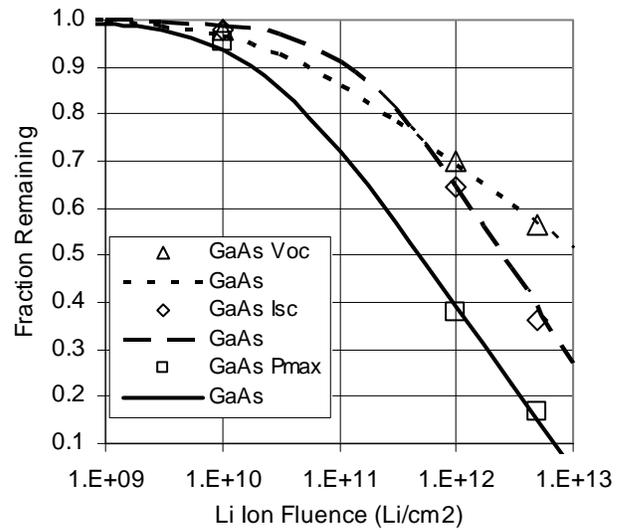

Fig. 5. IV parameter degradation rates (trendlines and data) of GaAs solar cells irradiated with 40 MeV Li ions.

Another case in Fig. 4 is the Isc value at $5 \times 10^{12}$. The AM0 curve for this fluence (in Fig. 1) indicated a series resistance problem. Therefore, the V = 0 point does not give the true Isc value and this point will fall off the trendline.

The only deviation from the trendline in Fig. 5 was in the Pmax value at $1 \times 10^{10}$. Since this level of radiation-damage is smaller than that of the other fluences, the percent error is larger (for beam fluence, IV measurements, and fitted curve). The standard degradation equation may not be valid in a region where the dominant defect is changing. We have chosen to maintain the form of the fitting curve (2) for consistency.

Fig. 6 compares the Pmax/Po values of the 15 and 40 MeV Li-ion-irradiated cells. The major difference is in the magnitude of the degradation rates. However, the Pmax values of the 15 MeV curves (with the exception of the $1 \times 10^{12}$ curve) are very close to those of the 40 MeV curves at a higher fluence. The same degradation values at Pmax/Po = 0.8 differ in fluence by approximately 35%. (It takes 35% more 40 MeV Li ions, compared to 15 MeV Li ions, to create the same degradation.) The same degradation values at Pmax/Po = 0.2 differ in fluence by approximately 40%. This effect is not much different from the near-equal damage for equal fluences of 15 and 40 MeV protons on 10 Ωcm silicon cells [7]. It is dramatically different than that predicted from the NIEL results, which predict that the fluence of 40 MeV Li ions should be 5 times that of the 15 MeV ions to produce the same degradation.

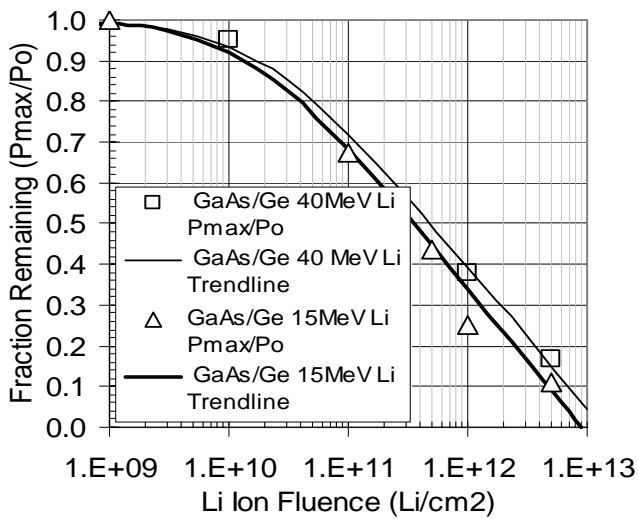

Fig. 6. Pmax/Po degradation rates (trendlines and data) of GaAs solar cells irradiated with 15 and 40 MeV Li ions.

## IV. DARK IV RESULTS

Figs. 7 and 8 display the dark-IV curves for unirradiated and irradiated silicon cells. Fig. 7 is for the unilluminated case and Fig. 8 is for the AM0 case. The AM0 dark-current characteristic is the IV curve with Isc subtracted. The advantage of this latter-type measurement is that it provides information under the operative illumination conditions.

The data in the two figures have not been corrected for cell area since this is a qualitative study and quantitative results are not needed, or useful, at present. Therefore, the results are shown in terms of the dark currents rather than the current densities. The data in Fig. 8 have been slightly smoothed to reduce the scatter observed in the low voltage region. (This fluctuation results from "flicker" (< 1%) in the solar simulator.)

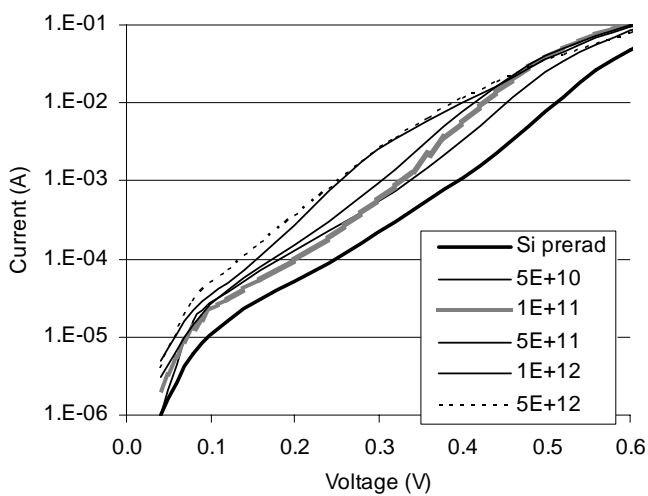

Fig. 7  Current-Voltage curves for unilluminated silicon solar cells with and without 40 MeV Li irradiation.

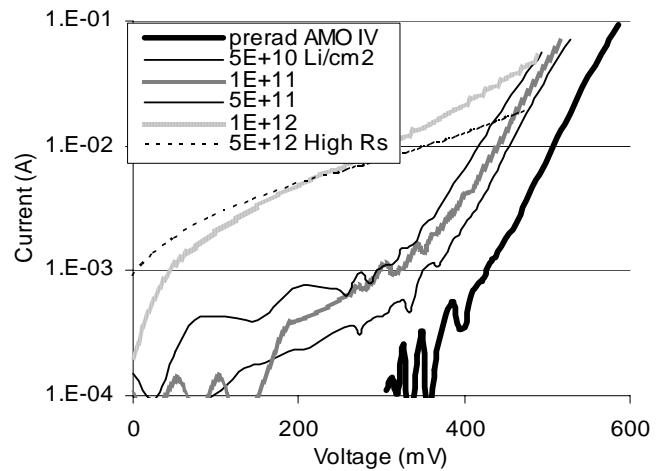

Fig. 8  Dark IV curves for AM0-illuminated silicon solar cells (Isc subtracted) with and without 40 MeV Li irradiation.

There are 3 major differences observed in the unilluminated and the AM0 dark currents. Firstly, the "noise" in the low dark-current data, as mentioned above, limits the AM0 dark-current plot to above 0.1 mA in Fig. 8. Secondly, the junction-shunt current (Vj/Rsh) is seen to be an order of magnitude higher for the irradiated cells in the AM0 case. (Compare dark currents for the $5 \times 10^{11}$ curves at 0.1 V.)

Thirdly, the series resistance affects the curves differently. In Fig. 7, the curves at high voltage (also high current) bend over, since the IRs term in this region contributes more to the difference in applied and junction voltage (Vj = V+ IRs, with negative current in (1)). In Fig. 8, maximum current flows at low voltages. If Rs is not small, one must correct V to obtain Vj since I is not zero when the applied voltage is zero.[b] This effect is most noticeable at the higher fluences ($1 \times 10^{12}$ and $5 \times 10^{12}$) in both figures.

All of the data and figures presented for the Li ion irradiation of silicon cells are typical of prior work with other radiation sources. The damage is primarily to the base region and the dominance of this effect is reflected in the IV curves.

The results for the unilluminated and the AM0 dark currents of GaAs cells, in Figs. 9 and 10, are qualitatively different from those of the Si cells (Figs. 7 and 8). The unilluminated (Fig. 9) and the AM0 curves (Fig. 10) for GaAs are quantitatively similar. While the effects of Rs are noticeable, in both Figs. 9 and 10, they are much less than those in the figures for Si (Figs. 7 and 8). The large separation (and large difference in shape) of the highest fluence curves in Fig. 8 (but not in Fig. 7) from the rest of the family of curves, is an artifact of the high radiation-induced series resistance.

The large difference in shape of the highest fluence curves in Figs. 9 and 10, from the rest of the family of curves, is a result of the radiation-induced damage, but not from an increase of series resistance. The nature of this defect type in GaAs is addressed below.

---

[b]  I ≠ Isc at V=0; but, I = Isc, when Vj = V+ Isc * Rs = 0. This effect is only significant under high illumination and for cells with high internal resistance.

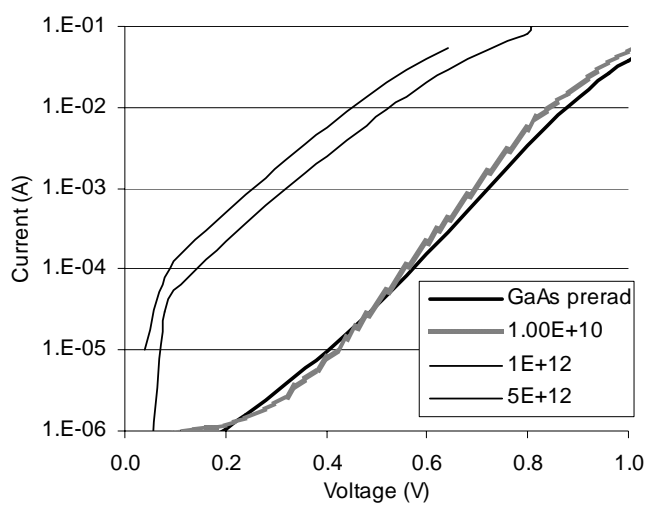

Fig. 9   Current-Voltage curves for unilluminated GaAs solar cells with and without 40 MeV Li irradiation.

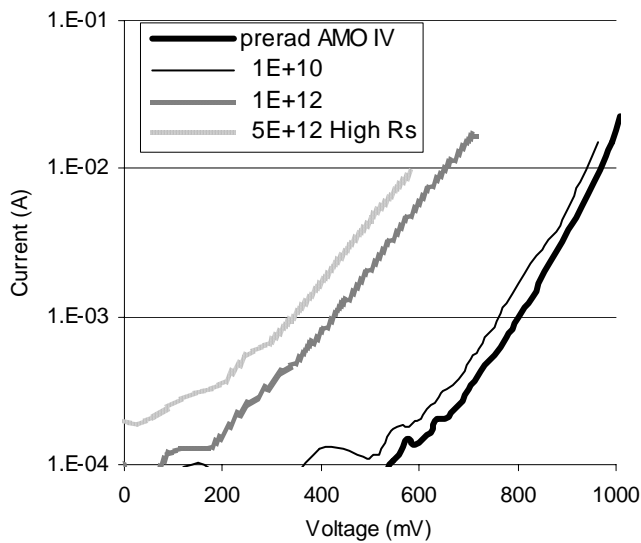

Fig. 10   Dark IV curves for AM0-illuminated GaAs solar cells (Isc subtracted) with and without 40 MeV Li irradiation.

## V. Discussion

We have provided a data set of lithium-ion-irradiated silicon and GaAs solar cells for comparison of present results with prior analysis of p irradiated Si and GaAs/Ge cells of similar types [8], [9]. Tables I and III provide the necessary information to compare present cells with those used in prior work. Tables IIIA and B provide uncorrected and cell-area-normalized values, respectively, for normalization of the relative data presented in the figures. Tables IIIB also includes Spectrolab silicon cell [10] values, which correspond better with the RWE cells of this work.

The K6 2 Ohm-cm Si cells from [8] are of a somewhat different variety from the 2 Ohm-cm RWE cells in the present work (and also from the presently available K6 cells, which are 10 Ohm-cm). However, the higher starting Isc, Voc, efficiency and Pmax of the K6 cells (resulting from the back-surface field) can be mathematically adjusted to make their degradation curves similar to those of the K4702 cells, after $1 \times 10^{11}$ p/cm$^2$. These adjusted proton data better represent the cell degradation in the present work than do those data for other readily-available proton-irradiated cells.

TABLE III:
UNIRRADIATED SOLAR CELL PERFORMANCE.
(IIIA: UNCORRECTED VALUES; IIIB: CELL-AREA CORRECTED VALUES.)

| (A) CELL TYPE | Isc (mA) | Voc (mV) | Pmax (mW) | Eff. (%) |
|---|---|---|---|---|
| Si------RWE | 1034 | 595 | 463 | 13 |
| Si [8] ------Spectrolab K6 | 342.4 | 608 | 157.6 | 14.6 |
| GaAs/Ge----Tecstar | 255 | 1020 | 194/200[c] | 17.5/18 |
| GaAs/Ge [8]----ASEC | 252 | 940 | 180 | 16.6 |
| GaAs [8]----ASEC | 252.8 | 996 | 202.4 | 18.7 |
| GaAs/Ge [11]----ASEC | 124.1 | 1007 | 97.9 | 18 |
| (B) CELL TYPE | Jsc (mA/cm$^2$) | Voc (mV) | Pmax (mW/cm$^2$) | Eff. (%) |
| Si------RWE | 39.4 | 595 | 17.6 | 13 |
| Si [5]------Spectrolab | 42.8 | 608 | 19.7 | 14.6 |
| Si [10]---Spectrolab K4702 | 39.2 | 585 | 18.0 | 13.3 |
| GaAs/Ge----Tecstar | 31.9 | 1020 | 24.3 | 17.5 |
| GaAs/Ge [8]----ASEC | 31.5 | 940 | 22.5 | 16.6 |
| GaAs [8]----ASEC | 31.6 | 996 | 22.3 | 18.7 |
| GaAs/Ge [11]----ASEC | 31.0 | 1007 | 24.5 | 18 |

In the case of the GaAs on Ge cells, the prior data used was from pre- [8] and post- [11][d] 1990 cells. The process for growing GaAs on Ge substrates was not yet well developed in 1987, so the earlier GaAs/Ge cells had lower voltages and fill factors than did the later cells of the same type. As a consequence, we have included data from a set of GaAs cells fabricated on the same line for comparison in Table III. These latter cells, in fact, better represent the cells of the present work than do the GaAs/Ge cells of that pre-1990 period.

Figs. 11 and 12 display the relative degradation of proton and Li irradiated silicon and GaAs solar cells as a function of the NIEL displacement-damage dose instead of the particle fluences (as in Figs. 4, 5 and 6). The NIEL values corresponding to the proton and Li fluences were simulated by SRIM 2003 [12].

The 40 MeV Li-irradiated Si cells from this work (RWE) are compared (in Fig. 11) with the 10 MeV proton irradiated K6 cells of [8], labeled K6702 in the figure, and with simulated 10 MeV proton degradation of K4702 cells that were electron irradiated [11] and then (using the appropriate NIEL values [3]) had their degradation curve converted to that of 10 MeV protons. When the starting values of Pmax for the K6 cells in [8] were lowered, to remove the effect of the BSF (and thus to convert the results toward K4702 cells), the results (not shown) shift over to and fall on the K4702 curve in Fig, 11

---

[c] The higher value represents the higher fill factor of the cells prior to removal of the coverslide.
[d] These cells were actually irradiated with 9.5 MeV protons.

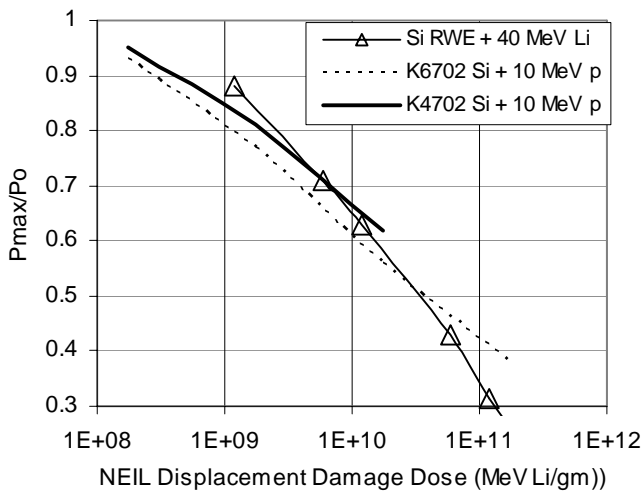

Fig. 11. Comparison of degradation rates for proton and Li irradiated Si solar cells in terms of displacement damage dose. The Li curve is from the Pmax/Po trendline of Fig.4, but plotted here in terms of displacement damage dose rather than fluence.

Since the RWE cells and the K4702 cells are very similar and the conversion, via NIEL process [3] to proton degradation results, is straight forward, these results should provide a good comparison of proton and Li-ion damage. The curves do overlap. However, they do not have the same slope. Therefore, it would appear that the nature of the damage is somewhat different for Li ions as compared with protons (or electrons). This difference is not obvious from the Si dark-IV characteristics of Figs. 7 and 8.

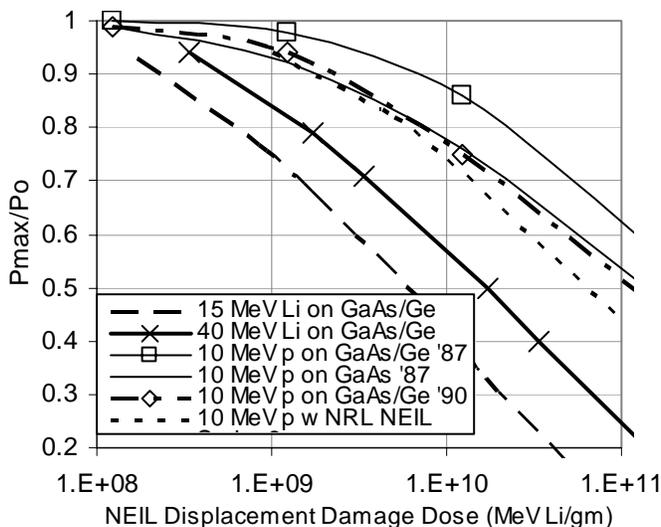

Fig. 12. Comparison of degradation rates for proton and Li irradiated GaAs cells in terms of displacement damage dose. The Li curves are from the corresponding Pmax/Po trendlines of Fig. 6, but plotted here in terms of displacement damage dose rather than fluence.

The 15 and 40 MeV Li-irradiated GaAs/Ge cells from this work are compared (in Fig. 12) with the 10 MeV proton irradiated GaAs/Ge and GaAs cells of [8] and with later GaAs/Ge cells [11]. The displacement damage doses, in all but the last curve, are determined from the proton and Li-ion fluences by the SRIM code values for NIEL. The last curve is taken from [3] for cell (GaAs/Ge, circa 1990) data from [11] and with NRL NIEL values from [3]. These same cell data, but with different SRIM 2003 NIEL values, are in fair, but not in exact agreement. (A difference in NIEL values is not unexpected, since the modeling of the appropriate choice is still in progress. However, the difference in slope between the last two curves, with the same cell data, is not understood.)

Clearly, within the present data and interpretation limitations, the NIEL process does not properly represent the Li ion damage to identical GaAs/Ge cells represented in the 15 and 40 MeV data of Fig. 12. Furthermore, while the process may equate displacement damage from gammas, electrons and protons [13], it does not equate the NIEL damage of all different radiation sources. We will examine a few possible explanations of this difference.

The first point is cell-type variation. The early GaAs/Ge cells (manufactured circa 1987) did not have as good a beginning-of-life (BOL) performance as later cells or those on GaAs substrates (~17% vs ~19% efficiency). This means that with radiation, the early cells will show less degradation until the radiation-induced damage dominates cell performance. Comparison of these two cell types in Fig. 12 shows the significant difference in results. A similar effect is seen between the K6702 and K4702 cells of Fig. 11. For this reason, the NIEL process is now predicated on the radiation damage from different sources being a relative value. Once the radiation performance of a cell type is determined for one radiation type, the calculated NIEL value of another radiation type allows prediction of the cell performance.

Knowledge of the cell type and its characteristics allows a non-arbitrary correction (rather than a normalization) to restore proper prediction using the NIEL process. As an example, if the proton irradiated '87 GaAs/Ge cells in Fig. 12 were given BOL values comparable to the GaAs cells (or later GaAs/Ge cells) in that figure, then the modified[e] degradation curve (not shown) is reduced from the original and falls onto that of the higher-efficiency cells. This indicates that the radiation-damage mechanism is the same in this class of cells and, when radiation damage dominates the IV characteristics, all of these cells will have similar values at high radiation levels.

In the case of the Li-irradiated GaAs cells, Po was reduced by the delidding process prior to irradiation. Adjustment for this loss (thus lowering Pmax/Po for all but the initial value), would lower both the 15 and the 40 MeV Li-irradiated-cell curves in Fig. 12. Such an adjustment would move the curves further from the predicted NIEL displacement damage values.

The second point has to do with the type of damage that radiation causes. Gammas, electrons, and protons, generally produce the majority of damage in the form of point defects that are uniformly distributed throughout a sample. A different type of damage that has been observed over the years [14],

---

[e] Since the trendlines in Figs. 4 – 6 are based on (2), an increase in adjusted Po will require a decrease in $\varphi_o$ to fit the adjusted data. This will shift the fitted curve downward.

[15] has not been dominant in any of the environments of interest. It *has* been of concern in laboratory simulation of space environments [14]. For this reason, the NIEL process was a welcome development, since testing with alternative sources (i.e., those less likely to cause anomalous damage) could be quantified, normalized, and accepted. It would appear that the use of Li ions has changed the dominant radiation-induced defect from point defects to cluster, or linear, defects.

## VI. Conclusion

A non-standard, energetic, radiation source (15 and 40 MeV Li ions) has been used to study radiation damage to Si and GaAs/Ge solar cells. The non-ionizing-energy-loss (NIEL) process has been successfully used to help in the comparison of these Li results with available proton-irradiation work. Analysis of the illuminated and dark IV characteristics has identified a different degradation mode (primary knock-on atom, PKA, related) in the GaAs cells compared to proton and electron degradation of similar cells. Plotting the Li and proton damage against the NIEL displacement damage dose (Figs. 11 and 12) has confirmed this analysis for GaAs/Ge cells and has indicated that this, or another, damage mode, while less dominant, may apply to silicon cells as well.

While this information may eliminate heavy ions as a "standard" simulation source, it indicates that Li-ion irradiation can be used as a tool for specific defect studies. Further analysis, of the cells from this work, and selected experiments are expected to lead to a better understanding of the damage mechanism(s) identified in the present work. Benefits of the NIEL process have been demonstrated in this paper. These have been achieved with the creation of appropriate proton-damage data sets for comparison with the present data), in the confirmation of unusual radiation defects in GaAs cells, and in the identification of a typical Si cell degradation mode. We hope that, in turn, the Li ion data provided here, and in future, will be useful in improving the range of applicability of the NIEL model and in reducing some of the limitations that have been observed.

## Acknowledgment

B. Jayashree is deeply indebted to the management, A.P.S Educational Trust, Bangalore and the University Grants Commission (UGC) New Delhi, India for allowing her to pursue research work in Bangalore University, India.

## References

[1] B.E. Anspaugh, *GaAs Solar Cell Radiation Handbook*, JPL Publication 82-69, 1982

[2] A. Meulenberg, "Basis for Equivalent Fluence Concept in Space Solar Cells," *Proceedings of the Silicon Research and Technology Workshop - Chairman's Report* NASA/LeRC, Cleveland, OH  4/82

[3] Insoo Jun; Xapsos, M.A.; Messenger, S.R.; Burke, E.A.; Walters, R.J.; Summers, G.P.; Jordan, T., "Proton nonionizing energy loss (NIEL) for device applications," IEEE Transactions on Nuclear Science,  50,  6  Dec. 2003 p. 1924

[4] A. Meulenberg, "Proton damage to Silicon from Laboratory and Space Sources,"  12th IEEE Photovoltaic Specialists Conf. (PVSC), Baton Rouge, LA,  November 15-18, 1976, Conf. Record

[5] B.E. Anspaugh, *GaAs Solar Cell Radiation Handbook*, JPL Publication 96-9, 1996.

[6] Inter-University Accelerator Center (IUAC), New Delhi, India http://www.nsc.ernet.in/infrastructure/accelerators/pelletron/index.htm

[7] A. Meulenberg and F. C. Treble, "Damage in Silicon Solar Cells from 2-155 MeV Protons," Conf. Record 10th IEEE Photovoltaic Specialists Conf. (PVSC), Palo Alto, Ca,  November 13-15, 1973, p.359

[8] R. H. Maurer, J. D. Kinnison, G. A. Herbert, and A. Meulenberg "Gallium Arsenide Solar Cell Radiation Damage Study" IEEE Trans. Nucl. Sci., vol. 36, pp. 2083-2091, Dec.1989.

[9] S. R. Messenger, G. P. Summers, E. A. Burke, M. A. Xapsos, R. J.Walters, E. M. Jackson, and B. D. Weaver, "Nonionizing energy loss (NIEL) for heavy ions," IEEE Trans. Nucl. Sci., vol. 46, pp. 1595–1602, Dec.1999.

[10] Spectrolab data sheets for Silicon K4702 Solar Cells http://www.spectrolab.com/DataSheets/K4702/k4702.pdf

[11] B.E. Anspaugh, "Proton and Electron Damage Coefficients for GaAs/Ge Solar Cells", Proceedings of the 22nd IEEE Photovoltaic Specialists Conference, Las Vegas, Nevada, p. 1593, October 1991.

[12] J. F. Ziegler, J. P. Biersack, U. Littmark, "The stopping and range of ions in materials," Pergamon press, New York, 1995.

[13] G. P. Summers, E. A. Burke, P. Shapiro, S. R. Messenger, and R. J.Walters, "Damage correlations in semiconductors exposed to gamma, electron and proton radiations," IEEE Trans. Nucl. Sci., vol. 40, pp.1372–1379, Dec. 1993.

[14] A. Meulenberg, "Proton damage to Silicon from Laboratory and Space Sources,"  12th IEEE Photovoltaic Specialists Conf. (PVSC), Baton Rouge, LA,  November 15-18, 1976, Conf. Record

[15] A. Meulenberg,  "Ionization-Induced Damage in Crystalline Silicon," Proc. of the High Efficiency and Rad. Damage Silicon Solar Cell Workshop NASA/LeRC, Cleveland, OH,  4/77 NASA p.221